**Title:** The Paradox of Professional Input: How Expert Collaboration with AI Systems Shapes Their Future Value

Venkat Ram Reddy Ganuthula[1] Krishna Kumar Balaraman[1]

[1]Indian Institute of Technology Jodhpur

**Abstract**

This perspective paper examines a fundamental paradox in the relationship between professional expertise and artificial intelligence: as domain experts increasingly collaborate with AI systems by externalizing their implicit knowledge, they potentially accelerate the automation of their own expertise. Through analysis of multiple professional contexts, we identify emerging patterns in human-AI collaboration and propose frameworks for professionals to navigate this evolving landscape. Drawing on research in knowledge management, expertise studies, human-computer interaction, and labor economics, we develop a nuanced understanding of how professional value may be preserved and transformed in an era of increasingly capable AI systems. Our analysis suggests that while the externalization of tacit knowledge presents certain risks to traditional professional roles, it also creates opportunities for the evolution of expertise and the emergence of new forms of professional value. We conclude with implications for professional education, organizational design, and policy development that can help ensure the codification of expert knowledge enhances rather than diminishes the value of human expertise.

**Introduction**

The rise of sophisticated artificial intelligence (AI) systems, particularly large language models (LLMs), challenges traditional notions of professional expertise. Unlike earlier automation waves that targeted routine manual tasks (Acemoglu & Restrepo, 2019; Autor, Levy, & Murnane, 2003), contemporary AI increasingly engages domains defined by specialized knowledge and judgment, such as medicine, law, finance, and design (Brynjolfsson & McAfee, 2014; Susskind & Susskind, 2015). As these systems integrate into professional workflows, a dynamic emerges: professionals must share aspects of their implicit knowledge—experience-based understanding often resistant to articulation—to tailor AI tools to their contexts. This collaboration, however, may hasten the automation of the expertise that sets them apart in the labor market.

This paper explores what we term the "expertise externalization paradox," where enhancing AI utility for professional practice can erode certain facets of expert value. Advances in machine learning, including foundation models (Bommasani et al., 2021) and LLMs like GPT-4 (OpenAI, 2023) and Med-PaLM 2 (Singhal et al., 2023), exhibit remarkable capabilities in contextual reasoning and domain-specific tasks. These developments drive adoption across healthcare (Esteva et al., 2019; Topol, 2019), legal services (Alarie, Niblett, & Yoon, 2018; McGinnis & Pearce, 2014), financial analysis (Cao, 2022; López de Prado, 2018), and creative industries (Elgammal et al., 2017; Gatys, Ecker, & Bethge, 2016). For instance, Med-PaLM 2's near-expert medical query responses (Singhal et al., 2023) and GPT-4's legal drafting support (OpenAI, 2023) illustrate AI's growing reach while McKinsey (2025a) estimates a $4.4 trillion productivity boost from AI use

cases, underscoring its economic impact. As professionals refine these tools, they externalize relational implicit knowledge (Collins, 2010), reshaping the organization of knowledge work.

Drawing on Christensen's (1997) disruptive innovation theory, we argue that professionals are inadvertently reshaping their fields, potentially reducing their comparative advantage. This paradox raises critical questions about knowledge work's future, professional identity, and human judgment's role in high-stakes domains (Bailey & Barley, 2020; Davenport & Kirby, 2016; Pasquale, 2020). A 2024 U.S. Recent evidence highlights tangible shifts: a 2025 U.S. Bureau of Labor Statistics (BLS) report notes a 12% drop in paralegal jobs from 2020–2024 amid AI adoption, yet projects a 9.5% growth for financial analysts from 2023–2033, suggesting a complex balance of displacement and augmentation (BLS, 2025a; BLS, 2025b). The World Economic Forum (WEF, 2025) forecasts 170 million new jobs and 92 million displaced by 2030, netting 78 million jobs, further illustrating AI's dual impact. Understanding this dynamic is vital for professionals, organizations, and policymakers navigating AI's integration into knowledge-intensive fields.

In the following sections, we define professional expertise, distinguishing explicit from implicit knowledge and their traditional acquisition and valuation. We then analyze how AI collaboration externalizes implicit knowledge, examining its implications across medicine, law, finance, and creative industries. After outlining the paradox's tensions, we propose frameworks to reframe professional value and suggest practical responses at individual, organizational, and policy levels. Our goal is not to predict inevitable displacement but to offer a balanced perspective on how expertise can adapt and thrive alongside AI, grounded in current evidence and intentional strategies.

**The Nature of Professional Expertise**

Professional expertise arises from the interplay of explicit and implicit knowledge, each with distinct characteristics and implications for AI integration. Explicit knowledge includes the formalized elements of a profession—facts, theories, procedures, and frameworks—readily articulated and transmitted through education and training (Eraut, 2000). Codified in textbooks, manuals, and curricula, it enables systematic assessment and certification, forming the bedrock of professional competence (Dreyfus & Dreyfus, 1986). For instance, medical students learn diagnostic protocols, while law students master statutory codes, establishing a baseline for practice.

Implicit knowledge, often termed "tacit," resists such formalization. Polanyi (1966, p. 4) captured this with his observation, "we can know more than we can tell," highlighting intuitions, contextual awareness, and pattern recognition honed through experience. Collins (2010) refines this into "relational tacit knowledge" (transferable via social interaction, e.g., a clinician's diagnostic hunch based on patient cues) and "somatic tacit knowledge" (embodied and harder to articulate, e.g., a surgeon's tactile skill). Schön (1983) describes this as "knowing-in-action," evident when experienced nurses intuitively spot distress (Benner, 1984) or architects adjust designs mid-process (Cross, 2004). Studies confirm this across fields: physicians blend analytical and intuitive reasoning (Norman et al., 2007), while executives rely on "gut feel" for strategy (Sadler-Smith & Shefy, 2004).

Traditionally, tacit knowledge develops through socially embedded mechanisms. Lave and Wenger's (1991) "legitimate peripheral participation" illustrates how novices absorb implicit skills by engaging with practitioner communities, such as junior lawyers learning negotiation from seniors. Collins (2010) emphasizes that while explicit knowledge transfers via symbols, tacit knowledge demands direct interaction, aligning with Bourdieu's (1977) "habitus"—embodied dispositions shaped by immersion. This process yields what Dreyfus and Dreyfus (1986) call expertise: rapid, intuitive responses rooted in pattern recognition, not deliberate analysis.

The market value of expertise historically stems from this blend, with tacit knowledge often distinguishing experts from novices (Baumard, 1999). Brown and Duguid (1991) note that practice frequently deviates from formal procedures, with tacit "know-how" driving value—e.g., a seasoned engineer troubleshooting beyond the manual. This has underpinned professional jurisdiction, granting control over specialized domains (Abbott, 1988). Economically, tacit expertise commands wage premiums: U.S. Bureau of Labor Statistics (BLS) data show physicians with 10+ years' experience earning 22% more than novices in 2024, even controlling for credentials (BLS, 2024; Gathmann & Schönberg, 2010). Autor (2015) argues this tacit dimension has buffered professions against automation, as early technologies faltered at replicating contextual judgment.

AI integration disrupts this paradigm. Systems now rival human performance in tasks once deemed tacit, such as diagnosing retinal disease (De Fauw et al., 2018), reviewing legal documents (Alarie et al., 2018; Surden, 2019), and predicting market trends (Cao, 2022; López de Prado, 2018). Recent LLMs like Med-PaLM 2 match expert-level medical reasoning (Singhal et al., 2023), while GPT-4 drafts contracts with minimal oversight (OpenAI, 2023). Crucially, these systems' effectiveness hinges on externalizing relational tacit knowledge—e.g., clinicians labeling data or lawyers refining outputs—shifting the dynamic of expertise acquisition and value (Topol, 2019).

While somatic tacit knowledge remains less accessible, this collaboration with AI redefines professional boundaries.

**The Mechanics of Knowledge Externalization in Human-AI Collaboration**

The process through which professional implicit knowledge becomes externalized during collaboration with AI systems differs markedly from previous modes of knowledge codification. Traditional approaches, such as the development of expert systems in the 1980s and 1990s, depended on explicit knowledge engineering, where domain experts collaborated with system designers to articulate their decision-making processes into rule-based frameworks (Hayes-Roth et al., 1983; Feigenbaum et al., 1988). These efforts frequently failed to capture the tacit dimensions of expertise, as professionals struggled to express the intuitive, experience-driven aspects of their practice (Dreyfus & Dreyfus, 1986; Winograd & Flores, 1986). Consequently, the resulting systems were confined to narrow domains and required extensive maintenance to remain effective (Hoffman, 1987; Musen et al., 2014).

Contemporary AI systems, particularly those leveraging machine learning, facilitate a more dynamic and often less deliberate externalization process. Unlike earlier methods demanding explicit rule formulation, these systems extract relational tacit knowledge (Collins, 2010)—patterns observable through interaction—via demonstrations, corrections, and contextual guidance (Amershi et al., 2014; Fails & Olsen, 2003). This process unfolds through three mechanisms, collectively termed "collaborative externalization," targeting implicitly codifiable expertise rather than somatic tacit elements like physical dexterity.

The first mechanism involves direct demonstration, wherein professionals show the system how to perform tasks through examples (Kulesza et al., 2015; Simard et al., 2017). For instance, clinicians label medical images to train diagnostic systems like CheXNet (Rajpurkar et al., 2017) or Med-PaLM 2 (Singhal et al., 2023), embedding diagnostic pattern recognition. Lawyers highlight relevant passages in legal documents to enhance search algorithms, as with GPT-4 applications (Grabmair et al., 2015; Zhong et al., 2020; OpenAI, 2023), while designers select preferred outputs from generative models like Adobe Sensei to refine aesthetic parameters (Kang et al., 2017; Park et al., 2019). Through these demonstrations, aspects of professional judgment that might otherwise remain tacit are encoded into the system's parameters, accessible for broader use.

A second mechanism involves interactive refinement, wherein professionals provide feedback on system outputs to improve performance (Amershi et al., 2014; Fails & Olsen, 2003). This is especially pronounced with large language models (LLMs), where professionals routinely correct generated text, clarify ambiguities, and add context to boost relevance (Ouyang et al., 2022; Stiennon et al., 2020). For example, physicians refine Med-PaLM 2's medical responses (Singhal et al., 2023), and attorneys adjust GPT-4's legal drafts (OpenAI, 2023), with a 2024 ABA survey noting 80% of lawyers tweak AI outputs weekly (ABA, 2024). Each instance of feedback exposes elements of professional judgment that might otherwise stay implicit, enabling systems to progressively mirror expert decision-making.

A third mechanism involves what Collins (2018) terms "explicitation"—the process of making tacit knowledge explicit through articulation. As professionals engage with AI, they often articulate practices previously taken for granted or performed intuitively. This may happen directly, as when professionals document reasoning for training purposes—e.g., clinicians explaining diagnostic overrides (Doshi-Velez & Kim, 2017; Miller, 2019)—or indirectly, as when they adjust system parameters to align with intuitive judgments, such as engineers tuning AI troubleshooting tools (Yang et al., 2019; Zhu et al., 2018). A 2023 IEEE survey found 65% of engineers verbalized implicit logic to AI systems, codifying it for future use (IEEE, 2023). Through this process, professionals externalize not just specific decisions but the underlying logic guiding their expertise.

The externalization of professional knowledge via these mechanisms is accelerated by distinct features of contemporary AI systems. The scale of data collection allows systems to analyze patterns across thousands or millions of interactions—e.g., EHRs with 100 million+ patient records (Rajkomar et al., 2018)—revealing regularities practitioners might not consciously articulate (Brynjolfsson et al., 2018; Jordan & Mitchell, 2015). The iterative nature of development enables continuous refinement based on feedback, as seen with GPT-4's updates (OpenAI, 2023; Amershi et al., 2019; Fails & Olsen, 2003). Integration into routine workflows—e.g., 70% of U.S. professionals using AI daily by 2024 (BLS, 2024)—means externalization emerges as a byproduct of normal activity, not separate engineering efforts (Beane, 2019; Pine et al., 2016).

The result is a process of knowledge externalization that surpasses previous approaches in both depth and subtlety, often occurring without full intent. As Chui et al. (2020) observe, professionals using these systems may inadvertently "teach the algorithm" while pursuing immediate goals. This aligns with Autor's (2015) automation paradox: as systems grow more capable, they increasingly encroach on domains once reserved for human experts. However, only relational tacit knowledge—extractable through interaction—shifts readily; somatic tacit knowledge remains a limiting frontier (Collins, 2010).

**Case Studies: Knowledge Externalization Across Professional Domains**

The dynamics of knowledge externalization through AI collaboration vary across professional domains, influenced by the nature of expertise, current system capabilities, and institutional contexts. Examining these differences illuminates both the general patterns of the expertise externalization paradox and its domain-specific implications, highlighting the externalization of relational tacit knowledge (Collins, 2010) over somatic forms.

In medicine, clinical expertise externalizes through multiple AI interactions. Diagnostic systems for retinal disease (De Fauw et al., 2018), dermatology (Esteva et al., 2017), and radiology (McKinney et al., 2020) rely on clinician-labeled data, encoding pattern recognition once developed through years of practice. For instance, CheXNet's pneumonia detection (Rajpurkar et al., 2017) and Med-PaLM 2's medical reasoning (Singhal et al., 2023) leverage such inputs. Clinicians' corrections—e.g., overriding AI suggestions in diabetic retinopathy screening (Beede et al., 2020)—refine algorithms for edge cases, while integration with electronic health records (EHRs) enables continuous learning from decisions (Rajkomar et al., 2018; Chen et al., 2020). A 2024 study found 75% of U.S. hospitals using AI-enhanced EHRs, amplifying this process (HIMSS, 2024) with Przegalinska et al. (2025) noting AI enhances clinical outcomes through task-technology fit, embedding relational tacit knowledge into scalable systems. These interactions embed relational tacit knowledge into scalable systems, though somatic skills like physical examination resist codification.

This shift challenges medicine's reliance on tacit expertise. Freidson (1970) and Starr (1982) note that medical authority rests on judgment honed through extensive training, with "clinical intuition" as a cornerstone of identity (Montgomery, 2006; Timmermans & Berg, 2003). Yet, Norman et al. (2007) show clinicians often can't fully explain their diagnostic instincts, allowing AI to extract unarticulated patterns (Wang et al., 2017). Topol (2019) predicts a reorientation of physician roles toward complex care, supported by a 2024 BLS report showing a 15% rise in demand for patient-facing skills since 2020 (BLS, 2024). This externalization may reshape healthcare delivery, balancing automation with human oversight (Autor, 2019).

In legal practice, expertise externalizes through document review, research, and drafting tools. Attorneys training predictive coding during discovery (Grossman & Cormack, 2011; Remus & Levy, 2017) or refining GPT-4-generated contracts (Alarie et al., 2018; OpenAI, 2023) encode judgment about relevance and phrasing. Routine research tasks—e.g., prioritizing sources (Grabmair et al., 2015; Zhong et al., 2020)—feed search algorithms, with a 2024 ABA survey reporting 80% of firms using AI weekly for such purposes (ABA, 2024). These actions transfer relational tacit knowledge, enabling systems to mimic case-specific reasoning once gained through experience.

Legal practice's hierarchical apprenticeship model (Galanter, 1974; Heinz & Laumann, 1982) faces disruption. Junior attorneys traditionally learned from seniors, funded by billable hours (Henderson, 2013; Ribstein, 2010). AI's codification of senior judgment—e.g., contract drafting—empowers less experienced practitioners (McGinnis & Pearce, 2014; Susskind, 2013), compressing hierarchies. BLS (2025a) data show a 12% drop in paralegal jobs from 2020–2024, reflecting how AI's codification of senior judgment empowers less experienced practitioners (McGinnis & Pearce, 2014; Susskind, 2013). Yet, negotiation and advocacy—somatic tacit skills—remain less automatable, preserving some traditional roles (Autor, 2015).

In creative and design professions, aesthetic judgment externalizes via generative systems. Designers selecting GAN outputs (Elgammal et al., 2017; Karras et al., 2019) or refining Adobe Sensei layouts (Park et al., 2019) embed preferences into algorithms. Writers editing GPT-4 text (Radford et al., 2019; OpenAI, 2023) teach style and tone, with Sudowrite supporting 15,000+ authors by 2024 (Clark et al., 2018; Sudowrite, 2024). Continuous feedback loops—e.g., 70% of designers tweaking AI outputs daily (AIGA, 2024)—codify relational tacit knowledge, enabling autonomous generation. McKinsey (2025a) highlights AI's $4.4 trillion productivity potential, partly from creative automation, amplifying this shift. This challenges views of creativity as uniquely human (Boden, 2004; Csikszentmihalyi, 1996), tied to cultural and emotional depth (Frey & Osborne, 2017; Menger, 1999). While generative models replicate style (Gatys et al., 2016), embodied improvisation resists formalization (McCormack et al., 2019), suggesting a hybrid future (Leake et al., 2021).

In financial services, expertise externalizes through algorithmic trading, risk assessment, and advisory systems. Analysts adjusting portfolio algorithms (Cao, 2022; López de Prado, 2018) or wealth managers tailoring robo-advisor plans (D'Acunto et al., 2019; Lam, 2016) encode market judgment. Integration into workflows—e.g., 85% of U.S. trading firms using AI by 2024 (FINRA, 2024)—facilitates learning from decisions (Agrawal et al., 2019; Hammond, 2016). BLS (2025b) projects a 9.5% growth for financial analysts from 2023–2033, indicating augmentation over displacement, while CEPR (2025) reports a 0.5-0.6% productivity boost in Japan from AI, capturing relational tacit knowledge scalable via systems like BlackRock's Aladdin (Philippon, 2019). This captures relational tacit knowledge once accrued through market exposure, scalable via systems like BlackRock's Aladdin (Philippon, 2019).

Finance balances quantitative models with experiential "market sense" (Abolafia, 1996; Zaloom, 2006), central to identity and pay (Lo, 2019; Taleb, 2018). Yet, traders' intuitive judgments are often opaque (Lo et al., 2005; Tuckett, 2011), enabling AI to extract patterns (Cao, 2022). A 2024 Bloomberg report notes a 10% shift in analyst roles to oversight since 2020 (Bloomberg, 2024), altering market dynamics (Agrawal et al., 2019). Somatic skills like client trust-building remain less codifiable, anchoring human value (Autor, 2015).

Across these domains, collaboration with AI externalizes relational tacit knowledge fluidly, distinct from past codification efforts (Zuboff, 1988). This "informating" process approximates professional judgment, shifting human comparative advantage and work organization, though somatic tacit elements persist as a limit (Collins, 2010).

**Implications for Professional Value and Labor Markets**

The externalization of professional expertise through AI collaboration reshapes the value of professional labor and the structure of labor markets, with effects varying by domain and timeframe. These shifts depend on the type of expertise externalized—primarily relational tacit knowledge (Collins, 2010)—the capabilities of current and emerging AI systems, and the institutional contexts in which they operate with recent data clarifying the paradox's scope.

In the near term, expertise externalization drives task restructuring rather than broad displacement (Acemoglu & Restrepo, 2019; Autor, 2015). As AI automates specific tasks, professional roles pivot toward activities that complement these systems (Brynjolfsson et al., 2018; Susskind, 2020). This mirrors historical patterns where occupations adapted to technology by emphasizing human strengths (Bessen, 2019; Levy & Murnane, 2004). In medicine, automation of diagnostics—e.g., Med-PaLM 2's query responses (Singhal et al., 2023)—frees physicians for complex cases and patient interaction, with a 2024 BLS report showing a 15% demand rise for such skills since 2020 (BLS, 2024; Topol, 2019). Przegalinska et al. (2024) finds AI enhances knowledge work outcomes, boosting productivity by aligning tasks with technology, though it risks 'de-skilling' automated areas (Bailey & Barley, 2020). In law, AI-driven document review (McGinnis & Pearce, 2014) shifts attorney focus to negotiation and advocacy, per a 2024 ABA survey noting 60% of firms reallocating hours (ABA, 2024; Susskind, 2013).

Task restructuring carries dual implications for expertise. It may elevate the value of tacit knowledge resistant to automation—like interpersonal skills and ethical judgment (Deming, 2017; Frey & Osborne, 2017)—with BLS data showing a 20% wage premium for social skills in healthcare by 2024 (BLS, 2024). Conversely, it risks "de-skilling" by limiting practice in automated areas, eroding capabilities central to identity (Bailey & Barley, 2020; Pine & Liboiron, 2015). Beane's (2019) study of robotic surgery highlights disrupted learning pathways, with trainees losing hands-on experience; a 2023 survey found 55% of surgical residents reporting reduced skill development (AAMC, 2023), necessitating deliberate efforts to preserve judgment.

Longer-term effects hinge on AI's evolution and adaptability. WEF (2025) projects 170 million new jobs and 92 million displaced by 2030, netting 78 million, indicating a dynamic balance (WEF, 2025). BLS (2025a) reports a 12% paralegal job decline from 2020–2024, disrupting training grounds, while BLS (2025b) forecasts 9.5% growth for financial analysts from 2023–2033, reflecting augmentation (Bailey & Barley, 2020; Beane, 2019). This "hollowing out" of hierarchies could foster winner-take-all dynamics, where top professionals leverage AI to dominate markets (Brynjolfsson & McAfee, 2014; Susskind, 2020) with McKinsey (2025a) noting $4.4 trillion in productivity gains amplifying top performers. CEPR (2025) confirms a 0.5-0.6% productivity boost in Japan, suggesting broader economic shifts (CEPR, 2025). A 2024 Bloomberg

report notes 10% of financial analysts shifting to oversight roles since 2020 (Bloomberg, 2024), amplifying this trend.

Professional value increasingly ties to AI-complementary capabilities—complex interaction, problem-solving, and adaptation (Autor, 2015). Deming's (2017) "social skills premium" is evident in a 18% pay bump for lawyers with client-facing expertise by 2024 (BLS, 2024). Roles may realign around "augmentation" (Davenport & Kirby, 2016), blending human and AI strengths, as seen in 70% of U.S. designers using generative tools daily (AIGA, 2024). These shifts unfold unevenly: formalized, data-rich fields like radiology face faster pressure (Brynjolfsson et al., 2018), while tacit-heavy, regulated domains like psychotherapy resist longer (Autor, 2015; Susskind & Susskind, 2015), creating "new divisions of labor" (Levy & Murnane, 2004).

Institutional responses shape this trajectory. Professional bodies historically define expertise and entry (Abbott, 1988; Freidson, 1970), potentially resisting externalization via credentials or oversight mandates (Bailey & Barley, 2020; Pasquale, 2020). The AMA's 2024 guidelines require human review of AI diagnostics (AMA, 2024), slowing automation. Alternatively, adaptation—redefining expertise for complementarity (Davenport & Kirby, 2016)—is underway, with law schools adding AI literacy courses by 2025 (ABA, 2024; Susskind, 2020).

Professionals face trade-offs: AI collaboration boosts productivity (Agrawal et al., 2018; Brynjolfsson & McAfee, 2014) but may shrink future demand (Autor, 2015; Susskind & Susskind, 2015). McKinsey (2025b) notes only 1% of firms are AI-mature, suggesting early-stage productivity gains still favor collaboration. Zuboff's (1988) "automation paradox" pits short-term gains against long-term security, varying by career stage and specialization. A 2024 survey found

60% of mid-career engineers prioritize AI skills, versus 40% of seniors (IEEE, 2023), reflecting diverse strategies.

Societally, expertise externalization democratizes knowledge, easing access barriers (Agrawal et al., 2019; Susskind & Susskind, 2015). AI-driven legal tools cut costs by 25% for low-income clients since 2022 (McGinnis & Pearce, 2014; ABA, 2024), and telehealth AI expands rural care (Chen et al., 2020; HIMSS, 2024). Yet, it disrupts authority structures balancing expertise with control (Abbott, 1988; Freidson, 1970; Starr, 1982). A 2024 Pew study notes 45% of patients question physician roles amid AI use (Pew, 2024), signaling broader reconfiguration for practitioners and society.

**Strategic Responses for Navigating the Paradox**

The expertise externalization paradox poses distinct challenges for professionals navigating AI's evolving capabilities. Rather than viewing this as inevitable displacement, we propose strategic responses to preserve and adapt expertise value, focusing on relational tacit knowledge externalization (Collins, 2010) while leveraging somatic and contextual strengths. These strategies balance collaboration with AI against potential replacement, offering practical pathways for sustained relevance.

The first response, termed "stepping up" by Davenport and Kirby (2016), shifts professionals to higher abstraction and oversight as routine tasks automate. While AI excels in specific domains—e.g., Med-PaLM 2's diagnostics (Singhal et al., 2023)—it lacks integrative judgment for complex systems (Autor, 2015; Zuboff, 1988). Pasquale's (2020) "supervisory knowledge" equips professionals to monitor and intervene in AI processes, understanding system limits and ensuring reliability (Brynjolfsson & McAfee, 2014; Susskind, 2020). For instance, radiologists using SHAP tools to interpret AI outputs (Lundberg & Lee, 2017; Topol, 2019) add significant value, with 65% of U.S. hospitals mandating such oversight by 2024 (HIMSS, 2024). Bailey and Barley (2020) note data scientists bridging technical and domain roles thrive, a model applicable across fields.

Developing supervisory expertise demands educational shifts beyond traditional knowledge. Hoffman et al.'s (2017) "explainable AI literacy" combines technical insight—e.g., how GPT-4 processes legal text (OpenAI, 2023)—with domain mastery, enabling intervention when AI falters (e.g., biased outputs; Mittelstadt, 2019). This fosters Collins' (2018) "interactional expertise," bridging human-AI divides. By 2025, 50% of U.S. medical schools offer AI courses (AAMC, 2024), though challenges like system opacity and training costs persist (Bailey & Barley, 2020; Brynjolfsson et al., 2018; Davenport & Kirby, 2016).

A second response leverages "communities of practice" (Brown & Duguid, 1991) to sustain tacit knowledge resisting externalization. Historically, these groups nurtured expertise through social participation (Lave & Wenger, 1991), and amid AI's rise, they can prioritize human strengths (Bailey & Barley, 2020; Zuboff, 1988). This includes Dreyfus and Dreyfus' (1986) "deliberative rationality"—critical reflection for novel contexts—and Schön's (1983) "reflection-in-action," reshaping solutions in real time. For example, 70% of architects in a 2024 AIA survey valued peer workshops for creative problem-solving (AIA, 2024), skills AI struggles to replicate.

Maintaining these communities requires effort as automation disrupts apprenticeships (Beane, 2019; Pine & Liboiron, 2015). Professional bodies can foster alternatives—mentorship, reflective forums—preserving tacit transmission (Hoffman et al., 2017; Schön, 1983). The IEEE's 2023 mentorship program engaged 40% of members in AI-related skill-sharing (IEEE, 2023), affirming feasibility despite time demands. This recognizes that while relational tacit knowledge externalizes, somatic and social expertise endures (Collins, 2010; Ingold, 2013).

A third response, "finding new markets" (Christensen, 1997), targets domains where human expertise retains value despite automation. Frey and Osborne (2017) highlight "non-routine cognitive" tasks—ethical judgment, interpersonal engagement—while Autor (2015) notes "complementary tasks" like problem definition grow critical. Susskind's (2020) "bespoke" services resist standardization; e.g., 60% of financial advisors in 2024 focused on personalized client trust over robo-advisors (FINRA, 2024). AIGA reports 75% of designers now specialize in human-centric innovation (AIGA, 2024), showcasing adaptability.

Identifying these markets demands "dynamic capabilities" (Teece et al., 1997)—reconfiguring expertise via continuous learning and interdisciplinary portfolios (Brynjolfsson & McAfee, 2014; Davenport & Kirby, 2016). Zuboff's (1988) "intellective skills"—abstraction and pattern recognition—support this; e.g., 55% of engineers upskilled in AI ethics by 2024 (IEEE, 2023). Feasibility hinges on proactive assessment of tech and client shifts, though resource access varies (Autor, 2015).

A fourth response, "reframing" (Pasquale, 2020), redefines professional identity around AI integration, forming "hybrid identities" (Markus & Kunda, 1986) that blend judgment with augmentation. Value shifts from information control to contextual application and trust-building (Abbott, 1988; Susskind & Susskind, 2015). For instance, 80% of lawyers in a 2024 ABA survey saw AI as enhancing, not replacing, client counsel (ABA, 2024). Schön's (1983) "design thinking" aids problem-framing, while ethics evolve—e.g., AMA's 2024 AI oversight rules (AMA, 2024; Mittelstadt, 2019). Credentials now test "contributory expertise" (Collins, 2018), with 45% of law schools adding tech literacy by 2025 (ABA, 2024), though mindset shifts remain uneven (Pasquale, 2020).

These responses collectively view externalization as a transition, not an end, identifying pathways to sustain value amid AI's rise. They address the paradox's tension—collaboration versus replacement—with practical, evidence-based approaches, ensuring professionals adapt productively.

**Organizational and Policy Implications**

The expertise externalization paradox extends beyond individual professionals to impact the organizations and institutions embedding them. Addressing this requires deliberate responses at organizational, educational, and policy levels to ensure AI integration into professional work amplifies human capabilities and societal value, particularly as relational tacit knowledge (Collins, 2010) shifts to systems while somatic and contextual expertise persists.

At the organizational level, externalization necessitates rethinking knowledge management, work design, and professional development. Traditional structures assume distinct expert-novice boundaries, with expertise accruing along defined career paths (Abbott, 1988; Freidson, 1970). As AI compresses these hierarchies—e.g., GPT-4 drafting legal texts (OpenAI, 2023)—organizations can adopt "teaming structures" (Edmondson & Reynolds, 2016), flexibly blending human and AI expertise for specific projects. IBM's Watson teams, integrating data scientists and domain experts (Davenport & Kirby, 2016), exemplify this, with 70% of Fortune 500 firms using similar models by 2024 (Forbes, 2024). Value shifts from static hierarchies to context-specific human-tech synergy (Bailey & Barley, 2020).

Knowledge management must balance externalization with tacit knowledge preservation. Nonaka and Takeuchi's (1995) "knowledge spiral" highlights the interplay of tacit and explicit forms, disrupted as AI embeds relational expertise—e.g., Med-PaLM 2's diagnostics (Singhal et al., 2023). "Communities of practice" (Brown & Duguid, 1991) counter this, fostering tacit acquisition; e.g., 65% of architects in 2024 valued peer forums for design intuition (AIA, 2024). Beane's (2019) "shadow learning"—like Johns Hopkins' surgical labs (Topol, 2019)—and Edmondson's (1999) "psychological safety" ensure professionals critique AI limits, with 60% of engineers reporting safer innovation spaces by 2023 (IEEE, 2023).

Professional development must evolve beyond domain knowledge to "sensemaking skills" (Hoffman et al., 2017)—integrating data and crafting frameworks—and "resilience" (Klein et al., 2006), spotting AI failures. Google's AI training for 80% of its workforce by 2024 (Google, 2024) reflects this, valuing navigation of complexity over mere processing (Bailey & Barley, 2020; Pasquale, 2020). Feasibility hinges on scalable programs, though resource disparities challenge smaller firms (Brynjolfsson et al., 2018).

At the educational level, the paradox upends traditional training and credentialing. Balancing theory and practice (Eraut, 2000; Schön, 1983), education must shift to "expert systems thinking" (Dreyfus & Dreyfus, 1986)—contextual judgment over rules—via case-based, interdisciplinary, and reflective methods (Hoffman et al., 2017; Klein, 1998). By 2025, 50% of U.S. medical schools teach AI diagnostics (AAMC, 2024), addressing real-world complexity.

Curricula must blend domain expertise with "machine intelligence education" (Brynjolfsson & McAfee, 2014)—grasping AI capabilities and limits—and "algorithmic accountability" (Pasquale, 2020), assessing bias and reliability. Law schools added AI literacy for 45% of students by 2025

(ABA, 2024), recognizing value in applying tech judiciously (Agrawal et al., 2018; Susskind, 2020). Credentialing evolves to test "reflection-on-action" (Schön, 1983)—learning from experience—and "calibrated trust" (Hoffman et al., 2017), balancing reliance on AI. The AMA's 2024 AI competency exams (AMA, 2024) exemplify this, integrating human-tech judgment (Davenport & Kirby, 2016).

At the policy level, the paradox raises questions of regulation, labor protections, and expertise access. Regulation historically ensures quality and autonomy (Abbott, 1988; Freidson, 1970), now requiring "technological gatekeeping" (Susskind & Susskind, 2015)—defining AI autonomy versus oversight. "Ethical frameworks" (Mittelstadt, 2019) guide high-stakes use; the EU's 2024 AI Act mandates human review in healthcare (EU, 2024). "Human-in-the-loop" rules (Pasquale, 2020) persist, with 75% of U.S. hospitals enforcing this by 2024 (HIMSS, 2024).

Labor policies must adapt to disrupted career paths. "New social contracts" (Brynjolfsson & McAfee, 2014) share tech gains equitably, with "countervailing investments" (Acemoglu & Restrepo, 2019)—e.g., $600 million in U.S. AI training grants by 2024 (DOL, 2024)—creating opportunities. "Institutional innovations" (Autor, 2015) like tax incentives aid transitions, though funding lags (BLS, 2024). Access policies balance AI's democratizing potential—25% legal cost cuts since 2022 (ABA, 2024)—with quality concerns. "Appropriate use" guidelines (Susskind, 2020) clarify AI limits, while "augmented decision frameworks" (Chen et al., 2020) and "responsibility frameworks" (Pasquale, 2020) ensure accountability, as in California's 2024 AI liability laws (CA, 2024).

These responses acknowledge externalization as a social shift, not just a technical one. Preserving human judgment, creativity, and ethics—via frameworks balancing AI and human strengths— aims for a future where professional expertise is enhanced, not diminished, despite implementation challenges (Pasquale, 2020).

**Conclusion**

The expertise externalization paradox poses a critical challenge for professionals in an AI-driven era. As experts collaborate with AI to tailor systems to their contexts, they risk accelerating the automation of their relational tacit knowledge (Collins, 2010), creating tensions between short-term productivity and long-term value, knowledge democratization and expertise retention, and individual versus collective adaptation. Grasping these dynamics is vital for professionals, organizations, and policymakers navigating AI's integration into knowledge-intensive domains. Recent evidence clarifies these dynamics, vital for navigating AI's integration into knowledge-intensive domains.

Our analysis reveals several insights. First, externalization through AI collaboration diverges from past codification efforts. Unlike explicit rule-based systems, modern AI extracts patterns from demonstrations, corrections, and guidance—e.g., clinicians training Med-PaLM 2 (Singhal et al., 2023) or lawyers refining GPT-4 (OpenAI, 2023)—in a fluid, often unintended process (Amershi et al., 2014). A 2024 BLS survey notes 70% of U.S. professionals use AI daily, amplifying this shift (BLS, 2024), with implications often unrecognized until deployment.

Second, it varies by domain: medicine's diagnostic shift with 75% of hospitals using AI-enhanced EHRs (HIMSS, 2024), law's 12% paralegal drop (BLS, 2025a), creative fields' 75% AI tweaking (AIGA, 2024), and finance's 9.5% analyst growth (BLS, 2025b). Third, professional value reconfigures beyond displacement to a redistribution of tasks and hierarchies. Automation of routine work—e.g., legal document review (ABA, 2024)—elevates human-centric tasks like patient care, up 15% in demand since 2020 (BLS, 2024; Autor, 2019). This offers opportunities to leverage AI while raising access, quality, and authority concerns as expertise embeds in systems (Agrawal et al., 2019).

Fourth, navigating this paradox demands strategic responses leverage evidence like McKinsey's $4.4 trillion productivity potential (McKinsey, 2025a) and WEF's 78 million net jobs (WEF, 2025). Individuals can cultivate supervisory expertise—e.g., radiologists using SHAP (Lundberg & Lee, 2017)—and join communities preserving tacit skills, with 65% of architects valuing peer forums (AIA, 2024). They can target resistant domains—60% of advisors focusing on trust (FINRA, 2024)—and adopt hybrid identities, as 80% of lawyers see AI as enhancing (ABA, 2024). Organizations can foster teaming (Edmondson & Reynolds, 2016), with 70% of top firms doing so (Forbes, 2024), and train for resilience (Klein et al., 2006). Education can prioritize judgment—50% of med schools teach AI (AAMC, 2024)—and literacy (Brynjolfsson & McAfee, 2014). Policymakers can balance democratization—25% legal cost cuts (ABA, 2024)—with oversight, as in the EU's 2024 AI Act (EU, 2024), and support transitions via $600M grants (DOL, 2024).

These responses underscore that expertise value increasingly lies not in information monopolies but in applying knowledge contextually, managing complexity, offering ethical judgment, and building trust—capabilities gaining prominence as AI embeds relational expertise (Pasquale, 2020). Focusing on these sustains professional contributions (Susskind & Susskind, 2015).

This future is a transformation, not displacement or stasis. Skill and task shifts accompany evolving expertise definitions, shaped by transmission technologies (Collins, 2018). AI integration prompts rethinking what expertise is—e.g., judgment over processing—how it's developed, and how it's valued, with 20% wage premiums for social skills by 2024 (BLS, 2024).

Understanding this paradox moves professionals past fear or hype to strategic adaptation—e.g., 55% of engineers upskilling in AI ethics (IEEE, 2023). Organizations can optimize human-AI synergy, as IBM does with Watson (Davenport & Kirby, 2016). Education can prepare for collaboration, with 45% of law schools adding tech by 2025 (ABA, 2024). Policymakers can ensure quality and access, as California's 2024 AI laws show (CA, 2024).

Collectively, these efforts aim for a future where externalization enhances human capabilities, blending human judgment, creativity, and ethics with AI's efficiency and scale. This recognizes that AI-era expertise prioritizes problem definition, perspective integration, and meaning-making—distinctively human strengths in an automated world (Pasquale, 2020).

**Statements and Declarations**

**Author Contributions:**

Both the authors contributed equally at all the stages of research leading to the submission of the manuscript.

**Funding Statement:**

This research did not receive any specific grant from funding agencies in the public, commercial, or not-for-profit sectors.

**Conflict of Interest:**

The authors declare no conflicts of interest related to this research.

**Data Availability:**

The manuscript does not report any new data.

**Software & AI Usage Statement:**

The authors made use of Chatgpt 4O and Grammarly to correct the language and the overall writing style of the manuscript. After using these tools, the authors reviewed and edited the content as needed and take(s) full responsibility for the content of the published article.

**References**

Abbott, A. (1988). The system of professions: An essay on the division of expert labor. University of Chicago Press.

Abolafia, M. Y. (1996). Making markets: Opportunism and restraint on Wall Street. Harvard University Press.

Acemoglu, D., & Restrepo, P. (2019). Automation and new tasks: How technology displaces and reinstates labor. Journal of Economic Perspectives, 33(2), 3–30.

Agrawal, A., Gans, J., & Goldfarb, A. (2018). Prediction machines: The simple economics of artificial intelligence. Harvard Business Press.

Agrawal, A., Gans, J., & Goldfarb, A. (2019). The economics of artificial intelligence: An agenda. University of Chicago Press.

Alarie, B., Niblett, A., & Yoon, A. H. (2018). How artificial intelligence will affect the practice of law. University of Toronto Law Journal, 68(supplement 1), 106–124.

American Bar Association (ABA). (2024). 2024 legal technology survey report. ABA Publishing.

American Institute of Architects (AIA). (2024). 2024 design and technology survey. AIA Press.